\documentclass[
reprint,
superscriptaddress,
 amsmath,amssymb,
 aps,
pra,
]{revtex4-2}

\usepackage{graphicx}% Include figure files
\usepackage{dcolumn}% Align table columns on decimal point
\usepackage{bm}% bold math
\usepackage{physunits}
\usepackage{xcolor}
\usepackage{amsmath}
\usepackage{natbib}

\begin{document}

\preprint{APS/123-QED}

\title{Spin current symmetries generated by GdFeCo ferrimagnet across its magnetisation compensation temperature}

\author{Héloïse Damas}
\email{heloise.damas@ru.nl}
\affiliation{Institut Jean Lamour, Université de Lorraine CNRS UMR 7198, Nancy, France}

\author{Michel Hehn}
\affiliation{Institut Jean Lamour, Université de Lorraine CNRS UMR 7198, Nancy, France}

\author{Juan-Carlos Rojas-Sánchez}
\email{juan-carlos.rojas-sanchez@univ-lorraine.fr}
\affiliation{Institut Jean Lamour, Université de Lorraine CNRS UMR 7198, Nancy, France}

\author{Sébastien Petit-Watelot}
\email{sebastien.petit@univ-lorraine.fr}
\affiliation{Institut Jean Lamour, Université de Lorraine CNRS UMR 7198, Nancy, France}

%\author{Jaafar Ghanbaja}
%\email{jaafar.ghanbaha@univ-lorraine.fr}
%\affiliation{Institut Jean Lamour, Université de Lorraine CNRS UMR 7198, Nancy, France}

\begin{abstract}
Ferrimagnets, composed of antiferromagnetically coupled magnetic sublattices whose net magnetisation can be tuned by temperature, offer a unique platform for probing the symmetry of the spin currents they generate and for identifying the sublattice contributions to these currents. Here, we investigate the spin current symmetries produced by GdFeCo ferrimagnet at a fixed concentration and across a broad temperature range, including the magnetisation compensation point. Using complementary techniques based on spin-torque ferromagnetic resonance spectroscopy, we separate the contributions of the spin Hall effect (SHE) and the spin anomalous Hall effect (SAHE). We show that the torques arising from both mechanisms retain their sign across the magnetisation compensation temperature, and that the SAHE-driven damping-like torque has the opposite sign to the SHE-driven term. We suggest that both effects originates from distinct electronic subsystems: the SHE emerging from Gd 5d electrons, and the SAHE from FeCo 3d electrons. Consequently, the SHE sign remains insensitive to the magnetisation state, whereas the SAHE sign does not invert at compensation, reproducing our observations. Together, these insights clarify the interplay between sublattices in ferrimagnetic spin transport and highlight the potential of ferrimagnetic spin currents to generate spin torques in adjacent layers or within the ferrimagnet itself.
\end{abstract}

\maketitle

\section{Introduction}
\begin{figure}[h]
  \centering
  \includegraphics[width=0.95\linewidth]{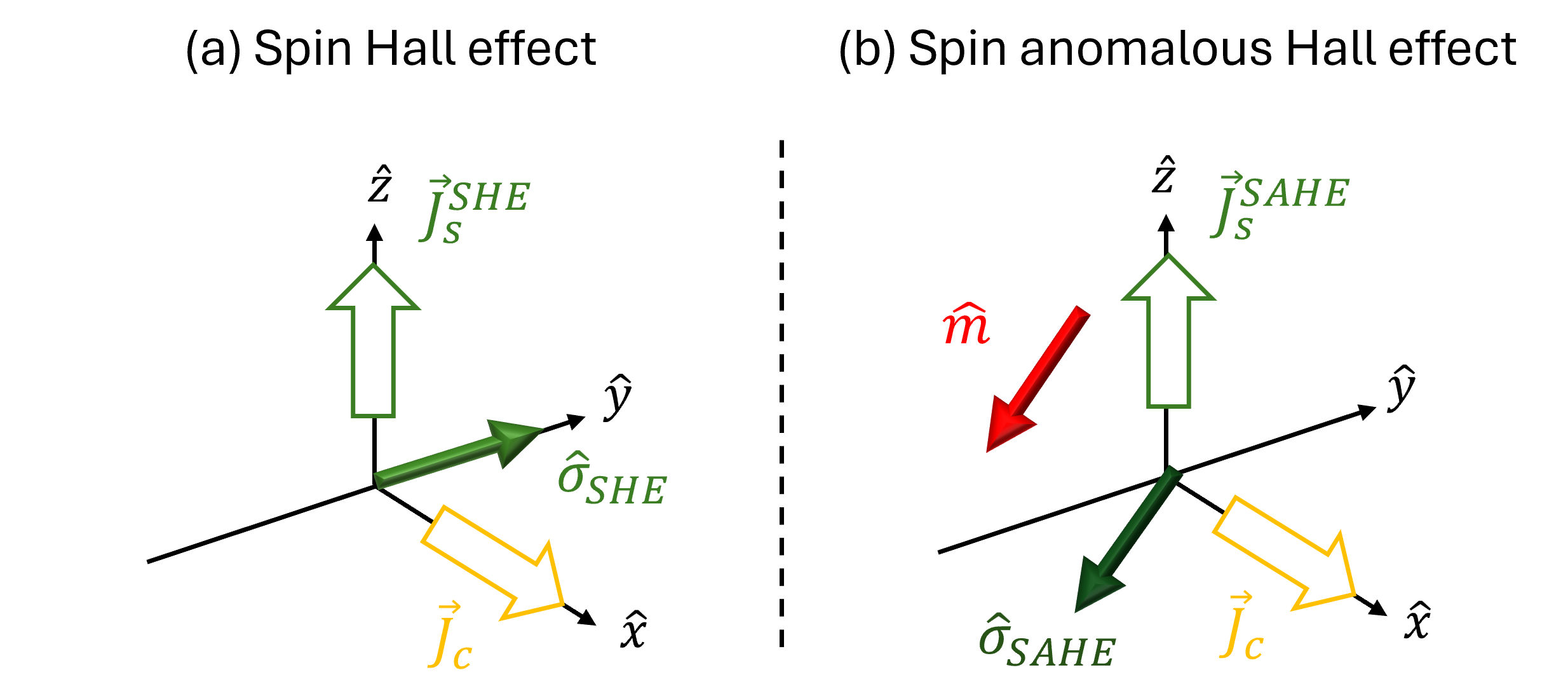}
  \caption{(a) Symmetry of the spin Hall effect (SHE) for the spin current $\vec{J}_s^{SHE}$ along the $\hat{z}$ axis. (b) Symmetry of the spin anomalous Hall effect (SAHE) for the spin current $\vec{J}_s^{SAHE}$ along the $\hat{z}$ axis.}
  \label{fig1}
\end{figure}

The spin Hall effect (SHE) is a well-established charge-to-spin conversion mechanism that typically occurs in non-magnetic heavy metals (HM) with strong spin-orbit coupling (SOC) \cite{hirsch1999spin, Nagaosa.2006,hoffmann2013spin, sinova2015spin}. In conventional systems, an applied charge current $\vec{J}_c$ is converted into a transverse pure spin current $\vec{J}_s^{SHE}$, with spin polarisation $\hat{\sigma}_{SHE}$ transverse to both the charge and the spin current directions. This relationship can be expressed as $\vec{J}_s^{\,SHE} = -\theta_{SHE} \, {\hbar}/{2e} \,  ( \hat{\sigma}_{SHE} \times \vec{J}_c )$ 
as illustrated in Fig.~\ref{fig1}(a). The resulting spin current can exert a spin-orbit torque (SOT) on adjacent magnetic layer \cite{Manchon.2009, Miron.2011, manchon2019current,shao2021roadmap}, enabling a wide range of spintronic applications such as magnetic random access memories (MRAM) \cite{prenat2015ultra,cubukcu2018ultra} and SOT nano-oscillators \cite{chen2016spin, zahedinejad2020two, zahedinejad2022memristive}.

Recent theoretical and experimental studies have revealed that magnetic materials can act as a source of spin currents with different symmetries that can generate a torque on an adjacent magnetic layer or even on themselves, giving rise to unconventional torques. The SHE, in particular, has been predicted and observed in magnetic systems \cite{amin2019intrinsic, zheng2024anatomy, zheng2025spin}, where the resulting spin current can exert torques either on adjacent magnetic layers \cite{wu2019spin} or on the magnetisation of the same layer, so called self-torque \cite{tang2020bulk, seki2021spin, ochoa2021self,cespedes2021current,aoki2023gigantic,ampuero2024self, kim2024spin}. Alongside the SHE, the anomalous Hall effect (AHE), traditionally used as a  probe of magnetisation \cite{Nagaosa.2010}, has a spin analogue known as the spin anomalous Hall effect (SAHE) \cite{Taniguchi.2015}. In the SAHE, a charge current is converted into a spin current according to $\vec{J}_s^{\, SAHE} = -\theta_{SAHE} \, {\hbar}/{2e} \,  (\hat{\sigma}_{SAHE} \times \vec{J}_c )$, where the spin polarisation $\hat{\sigma}_{SAHE}$ is parallel to the magnetisation, as shown in Fig.~\ref{fig1}(b). Such spin currents can generate torques in magnetic heterostructures with non-collinear magnetisations \cite{Iihama.2018, gibbons2018reorientable, baek2018spin, seki2019large}, or can even generate self-torques \cite{montoya2025anomalous}. Additional charge-to-spin conversion mechanisms have been identified in magnetic systems, including the planar Hall effect (PHE) \cite{safranski2019spin, safranski2020planar} and the magnetic SHE (m-SHE) \cite{mook2020origin, Wang2021, salemi2022theory}, both capable of producting unconventional SOTs. \\

Rare-earth transition-metal ferrimagnets (RE-TM FiMs) have emerged as promising sources of spin current, owing to the SOC of the RE element mediated by the intra-atomic ferromagnetic exchange interaction between the 4f localised electrons and the 5d conduction electrons \cite{watson1961exchange, campbell1972indirect}. In these materials, the magnetic moments of the RE and the TM are further coupled through an interatomic antiferromagnetic exchange interaction \cite{buschow1977intermetallic, brooks1991rare, duc1993exchange}, enabling the FiM properties to be tuned across both magnetic and angular momentum compensation points \cite{coey2010magnetism}. In FiM/HM bilayers, where the spin current is generated by HM and acts on FiM, it was experimentally demonstrated that the SOT sign remains unchanged across the magnetisation and angular compensation temperatures \cite{mishra2017anomalous, roschewsky2017spin}. However, when it comes to self-torque experiments on single-layer FiMs, where the FiM is the source of its own torque, discrepencies have been observed. For instance, no sign reversal of the self-torque is observed in single-layer GdFeCo \cite{krishnia2021spin}, a single reversal appears across the magnetisation compensation point in GdFeCo/Cu bilayers \cite{cespedes2021current}, and multiple reversals occur in FeTb across magnetic and angular compensation concentrations \cite{liu2022giant}. These observations raise fundamental questions. Do the observed sign changes originate from a sign change in the  generated spin current, or from changes in spin current absorption ? Is the spin current generation governed by the net magnetisation or by a specific magnetic sublattice ? Similarly, does the spin-current absorption couples to a particular sublattice or to the net magnetisation ? \\

In this study, we investigate the generation of spin currents with the SHE and SAHE symmetries by GdFeCo, maintaining the same ferrimagnet concentration while focusing on the spin current behaviour across the magnetisation compensation temperature. Using a GdFeCo/Cu/NiFe heterostructure, we probe the spin current generated by GdFeCo via spin-torque ferromagnetic resonance (ST-FMR) in NiFe. This approach allows us to disentangle contributions from the spin Hall effect (SHE) and the spin anomalous Hall effect (SAHE). Lineshape analysis isolates the SHE contribution, while DC-bias measurements capture the combined contributions of SHE and SAHE.
Near the magnetisation compensation temperature, the diverging anisotropy field in GdFeCo freezes the magnetisation out of the film plane, suppressing the SAHE contribution. Away from compensation, the magnetisation can be reoriented in-plane using an external magnetic field, enabling resolution of both SHE and SAHE contributions. By tracking the spin torques induced in NiFe across the ferrimagnet compensation, where one magnetic sublattice dominates over the other, we gain insight into the role of individual sublattices in the spin current generation.
Our results show that spin torques driven by both SHE and SAHE retain their sign across the magnetisation compensation temperature. Moreover, the damping-like torque driven by the SAHE exhibits an opposite sign and dominates over the SHE-induced torque at temperatures far from the compensations. This behaviour can be explained by considering that SHE and SAHE arise from different electronic sublattices: the 5d electrons of Gd and the 3d electrons of FeCo, respectively. These findings demonstrate that tuning the magnetic state of GdFeCo near compensation allows separation of spin currents with different symmetries, revealing the sublattice-specific origins of spin current generation.

\section{Sample and Measurement principle}

Measurements were carried out on a Gd$_{25}$Fe$_{65.6}$Co$_{9.4}$(10)/Cu(4)/Ni$_{81}$Fe$_{19}$(4)/Al(3) multilayer stack deposited on a Si/SiO$_2$ substrate by magnetron sputtering, where the numbers in parenthesis stand for the thickness in nanometers. The Cu spacer serves to magnetically decouple the two magnetic layers, while the Al capping layer prevents oxidation of the stack. Hall bar and spin–torque ferromagnetic resonance (ST-FMR) devices were patterned by photolithography, using identical current line dimensions of 50 $\mu$m in length and 10 $\mu$m in width. To ensure consistent magnetic properties, measurements were carried out on devices located in close proximity on the wafer.
\begin{figure}
    \centering
    \includegraphics[width=0.95\linewidth]{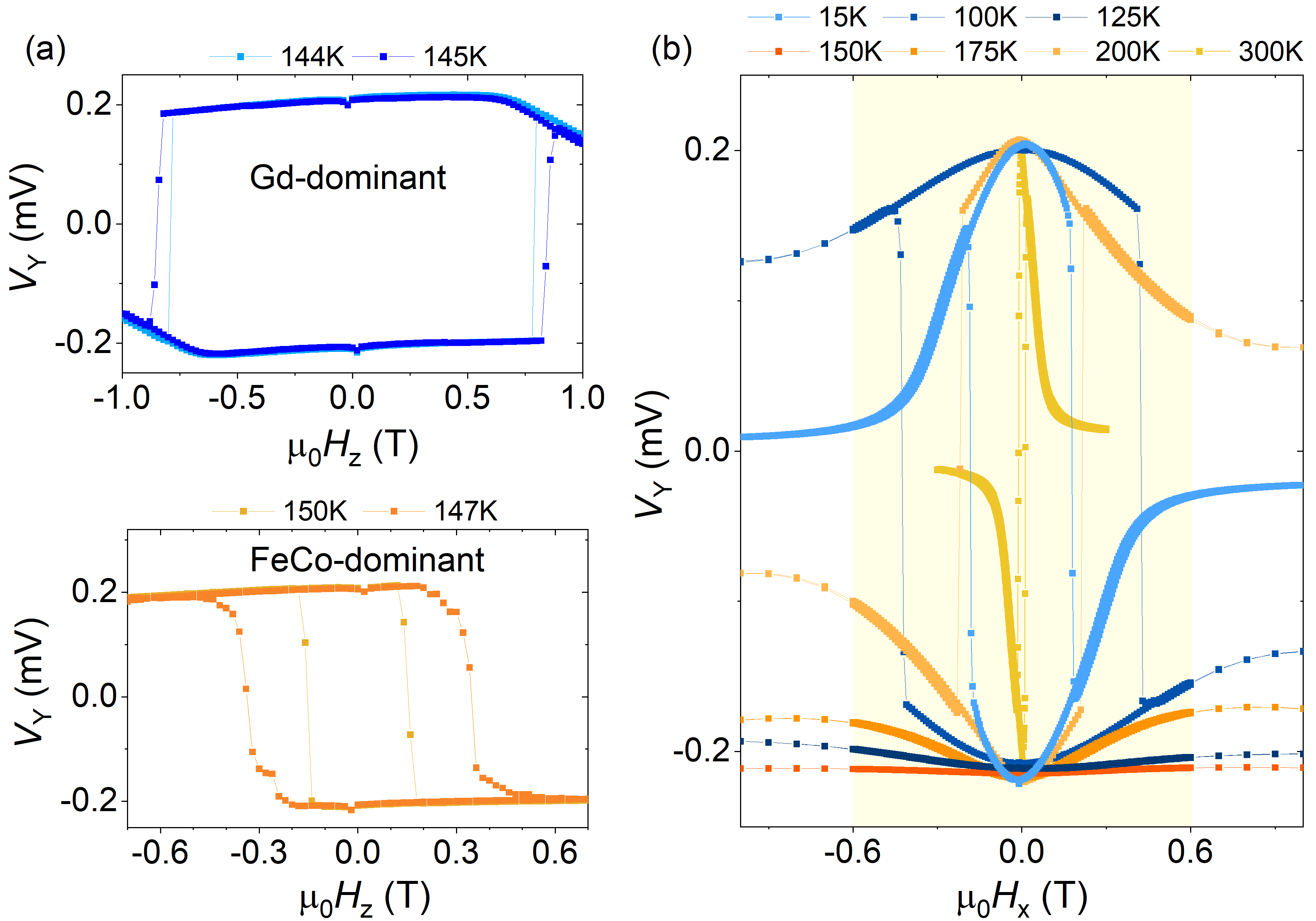}
    \includegraphics[width=0.95\linewidth]{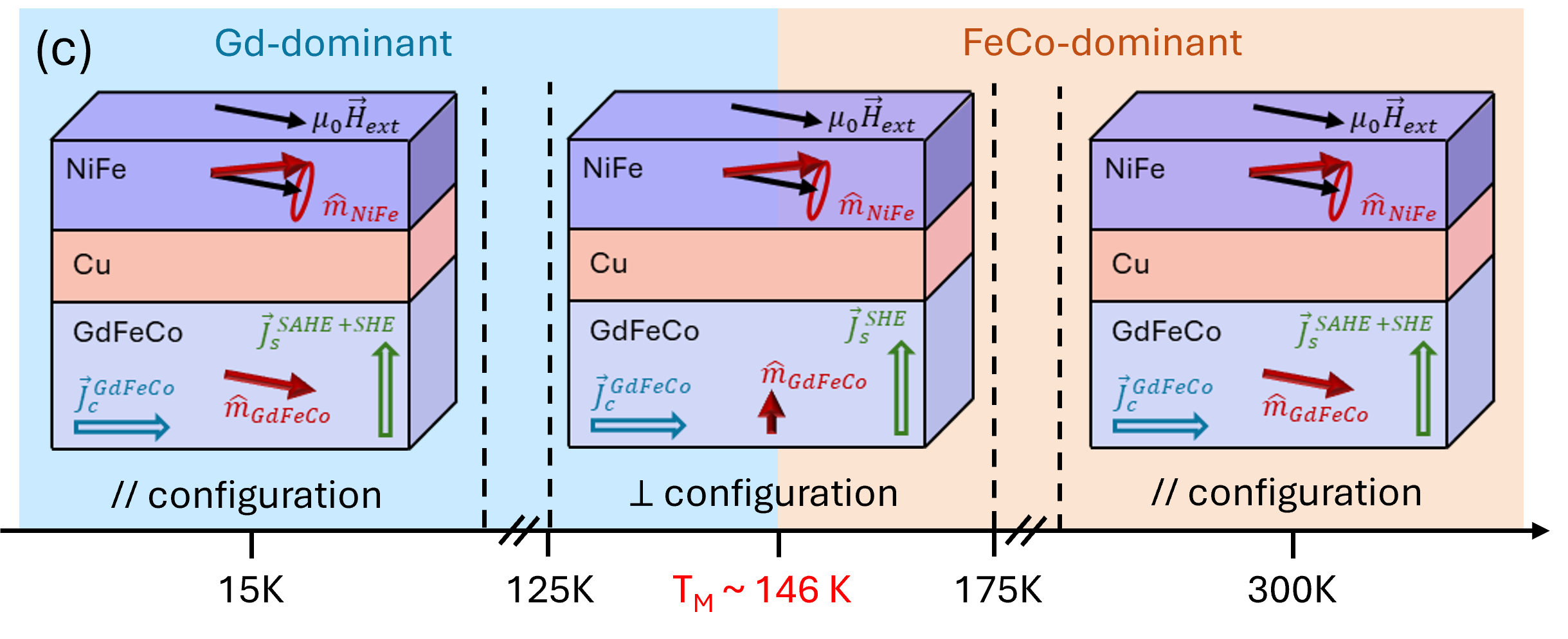}
\caption{(a) Hall voltage $V_Y$ as a function of an out-of-plane magnetic field for different temperatures. (b) Hall voltage $V_Y$ as a function of the magnetic field applied along the current line for different temperatures. The yellow background highlights the magnetic field range provided by the ST-FMR setup. (c) Schematic of the parallel ($\parallel$) and perpendicular ($\perp$) configurations for specific temperature ranges with the associated spin current symmetries.} 
\label{T charac}
\end{figure}

ST-FMR techniques are typically employed in FM/HM bilayers to probe spin currents generated in the HM and absorbed by the adjacent FM while its magnetisation is at resonance \cite{liu2011spin, wang2013ferromagnetic, wang2014determination}. In our multilayer heterostructure, both GdFeCo and NiFe layers exhibit distinct ferromagnetic resonances under different combinations of external magnetic field and microwave frequency. By exciting the resonance of the NiFe layer, we use it as a spin-current detector to probe the spin current emitted by GdFeCo, the latter being out of resonance.

The magnetisation compensation temperature ($T_M$) of GdFeCo was determined through magnetotransport measurements on Hall bar devices over the temperature range 15 K to 300 K. The root mean square current density set in these measurements was approximately $5.10^9$ A/m$^2$. The transverse voltage as a function of out-of-plane magnetic field is shown in Fig.~\ref{T charac}(a). A reversal in the voltage sign, corresponding to a sign change in the AHE across the magnetisation compensation, was used to estimate the compensation temperature as $T_M \approx 146$ K. The angular compensation temperature is estimated 30 K above $T_M$ \cite{haltz2020measurement, haltz2022quantitative}. The evolution of the transverse voltage with in-plane magnetic field shows the temperature dependence of GdFeCo anisotropy field. The yellow-shaded area in Fig.~\ref{T charac}(b) indicates the accessible field range in our ST-FMR setup. At temperatures far from compensation (e.g., 15 K and 300 K), the GdFeCo magnetisation can be aligned along the in-plane external magnetic field, while near compensation (e.g., 125 K to 175 K) it remains largely out-of-plane as the external field is not strong enough to overcome the anisotropy. These regimes define two distinct configurations for the ST-FMR measurements, as illustrated in Fig.~\ref{T charac}(c):
\begin{enumerate}
    \item \textbf{Parallel (//) configuration}: At temperatures far from $T_M$, both GdFeCo and NiFe magnetisations align with the in-plane external field. 
    \item \textbf{Perpendicular ($\perp$) configuration}: Near $T_M$, the GdFeCo magnetisation is oriented out-of-plane, while the NiFe magnetisation remains aligned with the external field in the plane. 
\end{enumerate}

Thus, temperature variation allows control of the magnetisation configurations, resulting in spin currents from GdFeCo with distinct symmetries. By analysing these spin currents under different configurations and employing complementary ST-FMR techniques (see Section \ref{section model}), the contributions from the various spin current symmetries can be disentangled. 

\section{Model}
\label{section model}
This section summarises the equations describing the spin torques and their corresponding effective fields arising from the symmetries of the SHE and the SAHE. It also outlines how different ST-FMR techniques can be employed to distinguish between these contributions.

\subsection{Spin torque symmetries}
We consider the case of the GdFeCo/Cu/NiFe heterostructure, in which the NiFe layer serves as a probe for the spin current generated by GdFeCo. The spin torque, $\vec{\Gamma}$, acting on the NiFe magnetisation $\hat{m}_{NiFe}$ ($\hat{v}$ denotes unit vectors) can be expressed, regardless of its origin, in terms of damping-like (DL) and field-like (FL) effective fields, $h_{DL}$ and $h_{FL}$:

\begin{align}
\dfrac{\vec{\Gamma}}{\gamma M_S^{NiFe}} &= \hat{m}_{NiFe} \times \left( h_{DL}\hat{\sigma}\times \hat{m}_{NiFe}\right)  + \hat{m}_{NiFe} \times h_{FL}\hat{\sigma},
\end{align}

where $\hat{\sigma}$ denotes the direction of the spin polarisation, $M_S^{NiFe}$ is the saturation magnetisation of NiFe and $\gamma$ is the gyromagnetic ratio of NiFe. The SHE symmetry is modelled by setting $\hat{\sigma}_{SHE} = \hat{y}$, while the SAHE symmetry is modelled as $\hat{\sigma}_{SAHE} = \hat{m}_{GdFeCo}$, where $\hat{m}_{GdFeCo}$ corresponds to the magnetisation direction of GdFeCo. For the two symmetries, the corresponding spin torques can be expressed as:

\begin{align}
\dfrac{\vec{\Gamma}_{SHE}}{\gamma M_S^{NiFe}} &= h^{SHE}_{DL} \hat{m}_{NiFe} \times \left( \hat{y}\times \hat{m}_{NiFe}\right) \nonumber \\ & + h^{SHE}_{FL} \hat{m}_{NiFe} \times \hat{y},
\end{align}
and 
\begin{align}
\dfrac{\vec{\Gamma}_{SAHE}}{\gamma M_S^{NiFe}} &= h^{SAHE}_{DL} \hat{m}_{NiFe} \times \left( \hat{m}_{GdFeCo} \times \hat{m}_{NiFe}\right) \nonumber \\ &+ h^{SAHE}_{FL} \hat{m}_{NiFe} \times \hat{m}_{GdFeCo}.
\end{align}

We define the total torque efficiency associated with the spin currents \cite{khvalkovskiy2013matching,Taniguchi.2015,Iihama.2018,manchon2019current, damas2022ferrimagnet}:
\begin{equation}
\xi_{DL,FL}^{SHE}= \dfrac{2e}{\hbar}  \dfrac{\mu_0 M_S^{NiFe}t_{NiFe}}{J_c^{tot}} h_{DL,FL}^{SHE},  
\end{equation}
and
\begin{equation}
\xi_{DL,FL}^{SAHE}= \dfrac{2e}{\hbar}  \dfrac{\mu_0 M_S^{NiFe}t_{NiFe}}{\left( \hat{m}_{GdFeCo} \times J_c^{tot} \right).\hat{z}} h_{DL,FL}^{SAHE},  
\end{equation}

where $t_{NiFe}$ is the thickness of the NiFe layer, and $h_{DL,FL}^{SHE,SAHE}$ is the SOT effective field. The term $\left(\hat{m}_{GdFeCo} \times J_c^{tot} \right).\hat{z} $ represents the spin current component emitted by GdFeCo along the $\hat{z}$ direction. When the magnetisation of GdFeCo is oriented out of the structure plane, the spin current with SAHE symmetry vanishes.

\subsection{Lineshape analysis}
\label{LS_FiM}
In the lineshape (LS) analysis experiment, the magnetisation dynamics of the spin-current probe layer is driven by both the radio-frequency (RF) Oersted field and the RF spin torque \cite{liu2011spin, wang2013ferromagnetic}. In the parallel configuration, the SAHE spin current can not induce a torque on NiFe magnetisation. In addition, in the perpendicular configuration, the SAHE spin current along the $\hat{z}$ axis vanishes. Consequently, the LS analysis isolates the influence of the SHE-induced spin current on the NiFe magnetisation. With this technique, the measured voltage can be decomposed into symmetric ($V_S$) and antisymmetric ($V_A$) components, which take the following forms when the anisotropic magnetoresistance (AMR) effect dominates the signal \cite{damas2022ferrimagnet}:

\begin{align}
V_{S}&=-\dfrac{i_{RF} R^{AMR}}{2} \dfrac{\sin{(2\varphi_H)} \cos{(\varphi_H})}{\mu_0(2H_{ext}+M_{eff}^{NiFe})} \dfrac{2 \pi f}{\gamma} \dfrac{h_{DL}^{SHE}}{\Delta_H^{NiFe} }, \nonumber \\ \noalign{\vspace{6pt}}
V_{A}&=- \dfrac{i_{RF} R^{AMR}}{2} \dfrac{\sin{(2\varphi_H)}\cos{(\varphi_H)}}{\mu_0(2H_{ext}+M_{eff}^{NiFe})} \dfrac{2 \pi f}{\gamma} \nonumber \\ & \hspace{3cm} \times \sqrt{1+\dfrac{M_{eff}^{NiFe}}{H_{r}^{NiFe}}}\dfrac{h_{FL'}^{SHE}}{\Delta_H^{NiFe}},
\label{VS,VA}
\end{align}

where $i_{RF}$ is the radio frequency current, $R^{AMR}$ is the amplitude of the AMR, $f$ is the frequency of the excitation, $H_{ext}$ is the in-plane external magnetic field, $M_{eff}^{NiFe}$ is the effective magnetisation, $H_r^{NiFe}$ is the resonance field and $\Delta_H^{NiFe}$ is the linewidth at resonance. $\varphi_H$ represents the angle between the applied current (DC and RF) and the applied magnetic field along which NiFe magnetisation equilibrium position is aligned. 
\noindent Here, the DL effective field $h_{DL}^{SHE}$ can be extracted from the symmetric part. The antisymmetric part provides access to the combined FL and Oersted contributions, expressed as $h_{FL'}^{SHE} = h_{FL}^{SHE} + h_{Oe}$. Because the RF current is challenging to extract, and because its amplitude relies on magnetoresistive effects, the LS analysis is not consider as a reliable method to quantitatively extract torque efficiencies. Here we will use this technique only to extract the signs of the SHE torque efficiencies.

\subsection{DC-bias}
\label{DC_FiM}
Applying a direct current (DC) during the ST-FMR measurements introduces a static torque on the oscillating magnetisation, thereby modifying the system’s dynamic susceptibility matrix \cite{Petit.2007, petit2008influence, ando2008electric}. This susceptibility originates from the magnetic energy density and is therefore linked to the effective field, $\mu_0 \vec{H}_{eff}$, along which the magnetisation aligns. Consequently, any component of the spin polarisation projected along this effective field can influence the NiFe magnetisation dynamics. Therefore, under a DC-bias, both the SHE and the SAHE can contribute. By extending theoretical frameworks previously developed for magnetic tunnel junctions \cite{Petit.2007, petit2008influence} and ferrimagnetic systems \cite{cespedes2021current, damas2022ferrimagnet}, the variation of the linewidth with respect to the applied DC current is expressed as:
\begin{align}
\dfrac{\partial \mu_0 \Delta_H^{NiFe}}{\partial i_{DC}} =& -\dfrac{ f}{ \gamma} \dfrac{1}{(2H_{r}^{NiFe}+M_{eff}^{NiFe})} \dfrac{1}{wt_{tot}} \nonumber \\ & \times \dfrac{\hbar}{2 \vert e \vert} \sin{(\varphi_H)} \dfrac{\xi_{DL}^{SAHE}+\xi_{DL}^{SHE}}{\mu_0 M_S^{NiFe} t_{NiFe}},
\label{DC_GFC_MD}
\end{align}
with $w$ the width of the current line and $t_{tot}$ is the thickness of the heterostucture conductive layers.\\

Similarly, the shift in resonance field induced by the DC current is given by:
\begin{align}
& \dfrac{\partial \mu_0 H_{r}^{NiFe}}{\partial i_{DC}}  =   \dfrac{\hbar}{2 \vert e \vert \omega t_{tot}} \sin{(\varphi_H)} \dfrac{\xi_{FL'}^{SAHE}+ \xi_{FL'}^{SHE}}{M_S^{NiFe} t_{NiFe}}.
\label{DC_GFC_FS}
\end{align}
This DC-bias technique therefore enables the detection of spin torques arising from both the SHE and SAHE, in contrast to the lineshape analysis, which is sensitive only to the SHE contribution. 

%\subsection{Remark on the LS analysis accuracy}
%Significant discrepancies have been observed between the torque measurements obtained using the LS technique and those derived from the DC-bias method. The LS technique relies on the decomposition of the measured voltage into symmetric and antisymmetric components, whose analytical expressions, given above, account only for the AMR. Applying these expressions in the presence of other magnetoresistive effects, such as the giant magnetoresistance (GMR), can lead to inaccurate determination of torque efficiencies. Additional contributions, discussed in Ref.~\cite{karimeddiny2021resolving} can also introduce residual signals in the LS analysis, further degrading its accuracy.

%In contrast, the DC-bias method does not depend on magnetoresistive effects. Instead, it determines the spin–torque efficiencies directly through induced modification of the magnetisation dynamics. Consequently, the DC-bias method is generally considered more reliable for accurately quantifying torque efficiencies. In the following, the LS analysis will therefore be employed only to discuss the signs of the spin torques, while their amplitude will be discussed based on the results obtained with the DC-bias technique.

\section{Results}
\label{section results}

The key parameters extracted from the LS and DC-bias analyses in GdFeCo(10)/Cu(4)/NiFe(4), which will be discussed in Section \ref{section discussion}, are summarised in Table \ref{Table_SAHE+SHE}. For comparison, results obtained for a reference Pt(5)/NiFe(4) bilayer are also included. The raw data and analysis can be found in the Supplemental Material (Appendix A to H).

\begin{figure}[h]
\centering
   \includegraphics[width=0.42\linewidth]{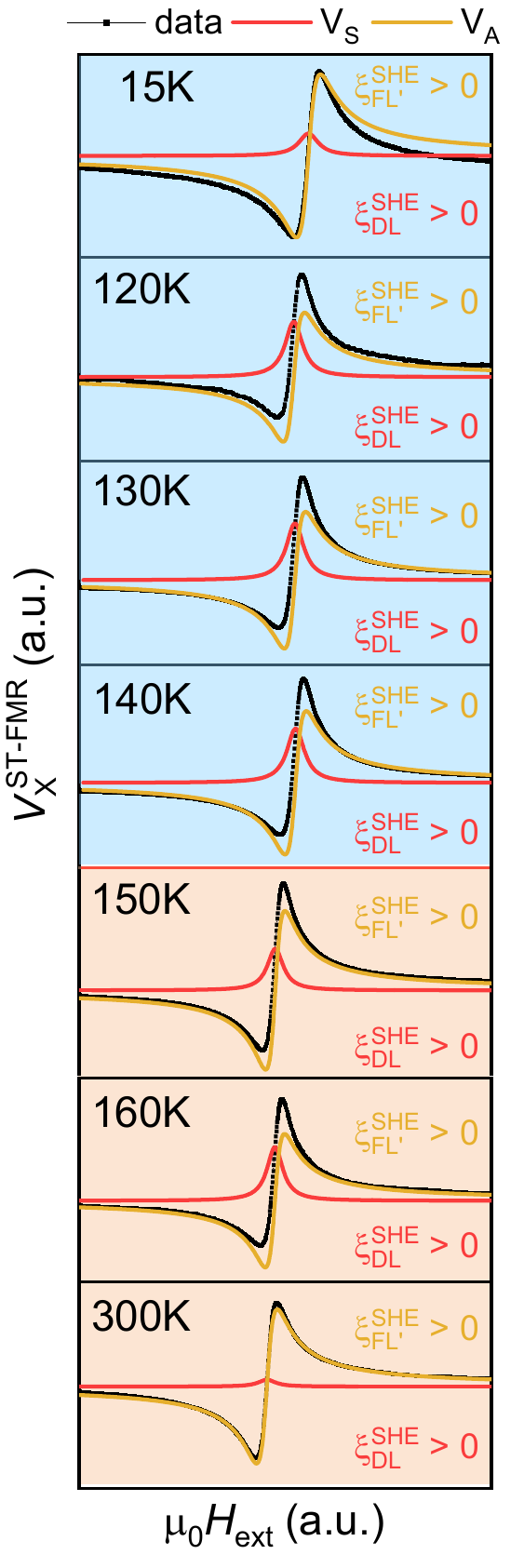}
  \caption{Summary of the lineshape analysis for temperatures ranging from 15 K to 300 K and for positive external magnetic field. The blue (orange) background corresponds to the Gd- (FeCo-) dominant regime. Across all temperatures, both the symmetric and antisymmetric components remain positive, i.e. the associated torque efficiencies are positive $\xi_{DL,FL'}^{SHE} > 0$.}
  \label{LS}
\end{figure}

\subsection{Lineshape analysis}

Representative ST-FMR spectra measured between 15 K and 300 K for positive applied magnetic field are shown in Fig.~\ref{LS}. Throughout the investigated temperature range, both the symmetric and antisymmetric voltage components are positive (negative) for a positive (negative) external magnetic field in a similar way as in our Pt/NiFe reference sample (see Appendix A). According to the model developed in Section~\ref{LS_FiM}, it results in positive efficiencies at all temperatures for both DL and FL' components, i.e. $\xi_{DL,FL'}^{SHE}$ is positive, suggesting that their signs do not change across the magnetisation and angular compensation temperatures.  %Based on Equation (\ref{VS,VA}), it can be inferred that the sign of the DL torque remains unchanged across the temperature range. The FL torque, however, is less certain since it is associated with a negative Oersted field, which may overshadow the FL torque and mask any potential sign changes. Therefore, the lineshape analysis indicates that the sign of the DL torque with the SHE symmetry acting on NiFe remains the same across the magnetisation compensation temperature of GdFeCo. 

\subsection{DC bias analysis}

\begin{figure}[h]
  \centering
  \includegraphics[width=0.9\linewidth]{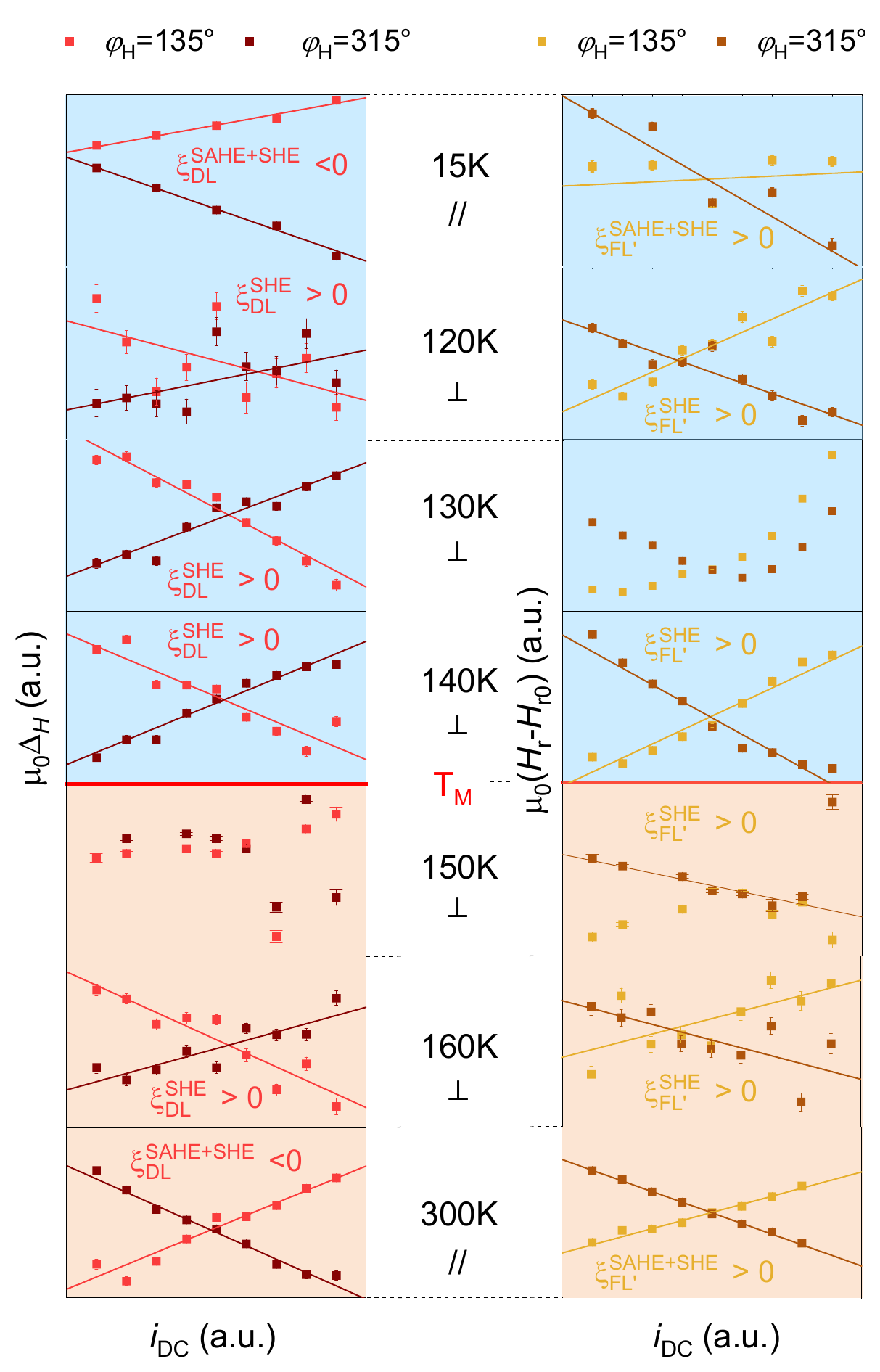}
  \caption{ Summary of the DC-bias analysis from 15 K to 300 K. The blue (orange) background corresponds to the Gd- (FeCo-) dominant regime. The left panel shows the linewidth modulation upon applied DC current, while the right panel shows the resonance field shift. In the perpendicular configuration ($\perp$), $\xi^{SHE}_{DL} > 0$ and $\xi^{SHE}_{FL'}>0$. In the parallel configuration ($\parallel$), $\xi^{SHE+SAHE}_{DL} < 0$ and $\xi^{SHE+SAHE}_{FL'} > 0$. The SHE and the SAHE symmetries have the same sign across the magnetisation compensation.}
  \label{DC biais}
\end{figure}

The results of the DC-bias analysis performed between 15 K and 300 K are summarised in Fig.~\ref{DC biais}. The left-hand panel presents the current-induced modulation of the linewidth, corresponding to the DL torque, while the right-hand panel shows the current-induced shift in resonance field, associated with the FL torque and the Oersted contribution. In the perpendicular configuration (120 K to 160 K), where only the SHE symmetry contributes to the spin current, both $\xi_{DL}^{SHE}$ and $\xi_{FL'}^{SHE}$ are positive, consistent with the LS results. In the parallel configuration (at 15 K and 300 K), where the spin current exhibits combined SHE and SAHE symmetries, $\xi_{FL'}^{SHE+SAHE}$ is also positive, while $\xi_{DL}^{SHE+SAHE}$ is negative. At 15K and 300K, the sign of $\xi_{DL}^{SHE+SAHE}$ remains negative, suggesting that the sign does not change across the compensation temperatures. %Since the sign of the SHE+SAHE DL torque is opposite to that of the SAHE DL torque alone, it is inferred that the SAHE contribution dominates over the SHE contribution. 
 
\section{Discussion}
\label{section discussion}
In this section, we first discuss the signs of the extracted torque efficiencies and provide a qualitative explanation for the absence of sign reversal in the SAHE and SHE symmetries across the magnetisation compensation temperature. We then discuss the amplitude of the torque efficiencies.  

\begin{figure}
  \centering
  \includegraphics[width=0.7\linewidth]{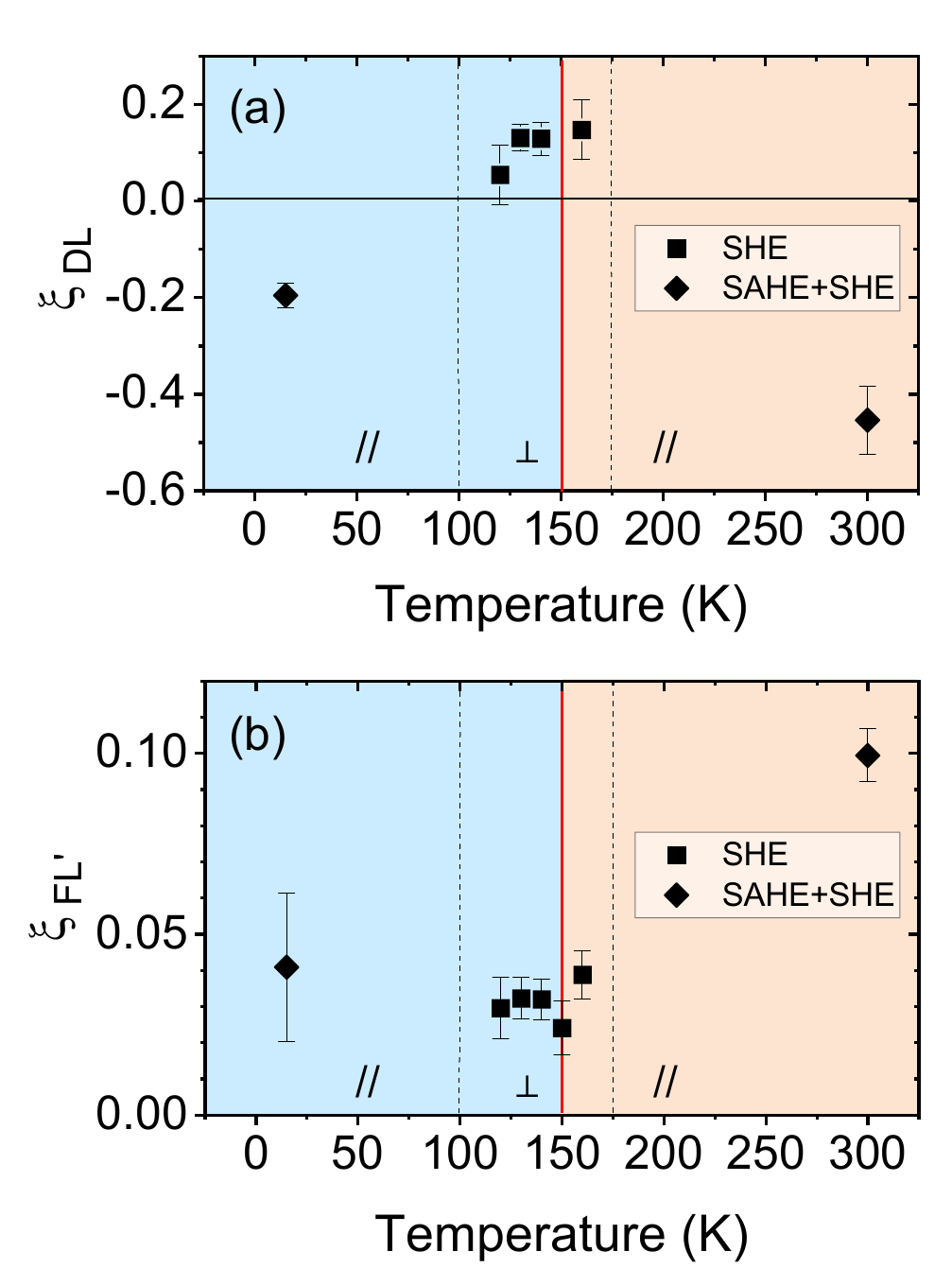}
  \caption{(a) DL and (b) FL' torque efficiencies extracted by the DC-bias technique as a function of temperature. The blue (orange) background corresponds to the Gd- (FeCo-) dominant regime. The dashed parallel lines at 100 K and 175 K separate the $\parallel$ and $\perp$ magnetisation configurations.}
  \label{summary}
\end{figure}

\subsection{Discussion on the efficiencies sign}

Fig.~\ref{summary} presents the DL and FL' torque efficiencies extracted from the DC-bias analysis. Consistent with the LS analysis, both $\xi^{{SHE}}_{DL}$ and $\xi^{{SHE}}_{FL'}$ remain positive across the magnetisation compensation point. In contrast, DC-bias measurements performed well outside the compensation region in the parallel configuration yield $\xi^{SAHE+SHE}_{FL'}>0$ while $\xi^{SAHE+SHE}_{DL}<0$, indicating that the SAHE contribution to the DL torque is opposite in sign to that of the SHE and dominant in this regime. Importantly, the SAHE sign also remains unchanged across both the magnetisation and the angular compensation points. We also note that the Oersted field is negative, therefore $\xi_{FL'}>0$ suggests that the FL torque is positive and dominates over the Oersted field. In the following, we aim to qualitatively explain the sign and the absence of sign reversal of the SHE- and SAHE- driven DL torques. Although both spin current symmetries can generate DL and FL torques, the DL torque is mainly due to the dephasing of the spin current inside the magnetic layer, whereas the FL torque is usually attributed to interfacial contributions \cite{Stiles.2002,haney2013current, mahfouzi2020microscopic}. For this reason, our qualitative model focuses on the DL torque.  \\

To begin our explanation, note that experiments on elemental Gd and FeCo films have shown that they exhibit anomalous Hall effects (AHE) of opposite signs \cite{chrobok1977spin} (see Appendix I). It suggests that the two elements exhibit opposite sign of spin-orbit coupling coefficient $\lambda_{SOC}$.

\begin{figure}
    \centering
    \includegraphics[width=0.75\linewidth]{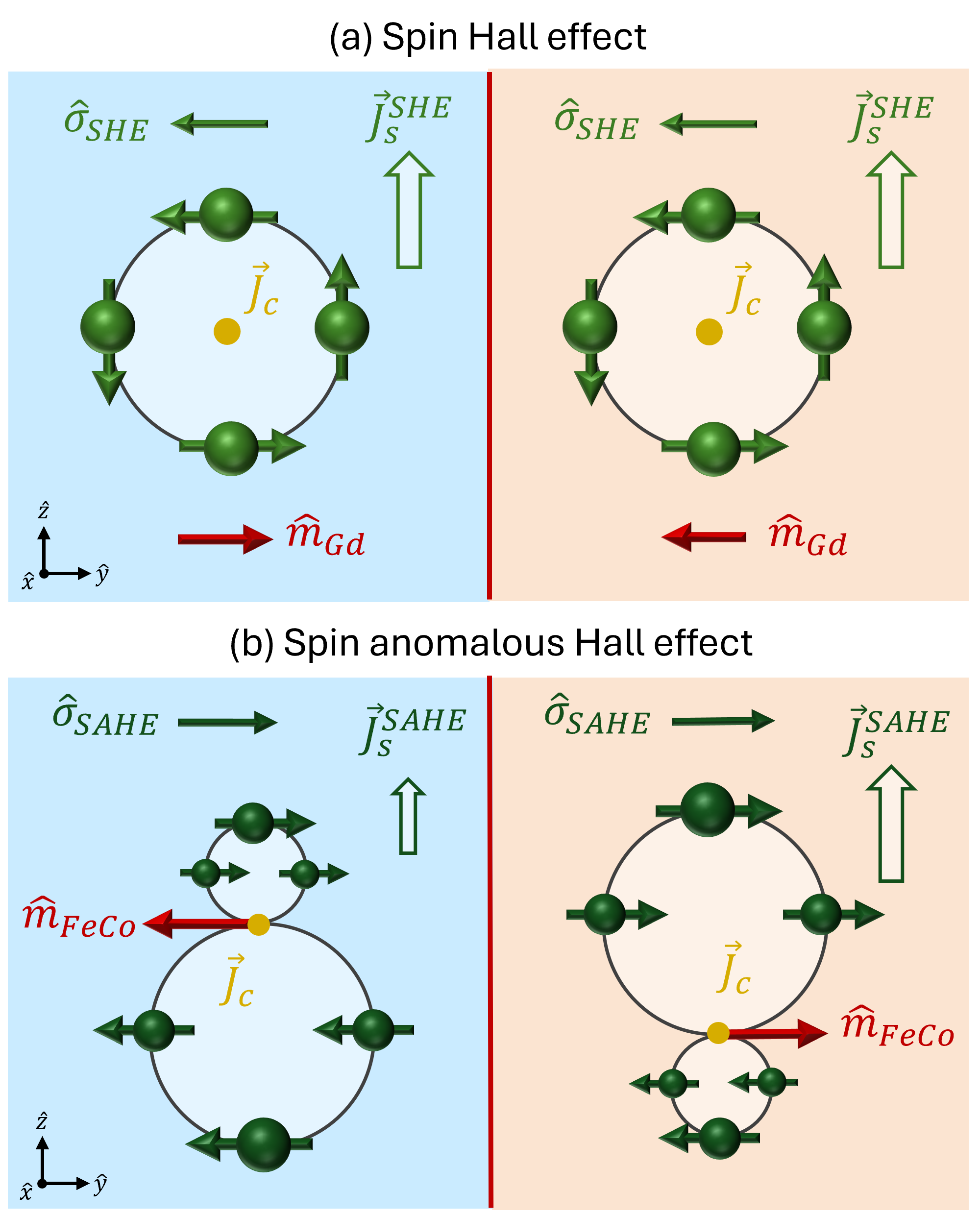}
    \caption{Schematic of the electron conduction for the (a) spin Hall effect and the (b) spin anomalous Hall effect symmetries, with an applied magnetic field along $+\hat{y}$. Assuming that these effects are caused by conduction bands with opposite sign of $\lambda_{SOC}$, we can explain why the SHE- and SAHE- driven torques have opposite signs. We can also see that the SAHE spin current has the same spin polarisation across the magnetisation compensation, explaining why the DL torque efficiency retains the same sign. The blue (orange) background corresponds to the Gd- (FeCo-) dominant region.}
    \label{fig:model}
\end{figure}

The SHE in nonmagnetic or magnetic metals arises from unpolarised electron bands. Therefore, in GdFeCo alloys, the SHE-driven spin current is attributed to the weakly polarised 5d band of the Gd sublattice. Fig.~\ref{fig:model}(a) illustrates the pure spin transport in this case. Because the SHE-driven spin current is generated by unpolarised states, its sign does not depend on the ferrimagnetic configuration but depends on the sign of $\lambda_{SOC}$ and therefore remains unchanged across the magnetisation compensation. This behaviour is consistent with theoretical predictions that the intrinsic SHE in magnetic systems with strong SOC is independent of magnetisation \cite{amin2019intrinsic}.

In contrast, the SAHE originates from spin-polarised conduction bands. In GdFeCo, this mechanism is expected to be dominated by the 3d electrons of the FeCo sublattice. Fig.~\ref{fig:model}(b) sketches the corresponding spin-polarised transport. Upon crossing the magnetisation compensation, the charge-carrier imbalance responsible for the AHE reverses, producing the well-known AHE sign change. However, the spin polarisation of the relevant carriers remains of the same sign as it is related to the sign of $\lambda_{SOC}$. This explains why the SAHE-related torque symmetry does not change sign across compensation.  

Taken together, these observations support the view that the SHE and SAHE originate from distinct electronic subsystems, namely the 5d electrons of Gd and the 3d electrons of FeCo, respectively. Because these channels possess opposite sign of $\lambda_{SOC}$ (as demonstrated by the opposite sign of AHE), the corresponding torque efficiencies also appear with opposite signs, as observed. The SHE is insensitive to the magnetisation state, while the spin polarisation relevant to the SAHE does not invert at compensation, explaining the absence of sign reversals.

An open question remains regarding the relative sign of $\lambda_{SOC}$ in Gd and FeCo compared with Pt. Hund’s third rule predicts that $\lambda_{SOC}$ becomes negative (positive) when the occupancy of the d conduction band is above (below) half-filling \cite{morota2011indication}. Following this expectation, one would anticipate $\lambda_{SOC} > 0$ for FeCo and Pt, and $\lambda_{SOC}  < 0$ for Gd \cite{sala2022giant}. Our experimental findings, however, indicate that Gd in Gd$_{25}$Fe$_{65.6}$Co$_{9.4}$ and Pt exhibit the same sign of $\lambda_{SOC}$. This discrepancy raises fundamental questions about how $\lambda_{SOC}$ signs originating from different sublattices combine within an alloy. In our work, the relative concentration of Gd and FeCo is kept constant while the temperature was changed, making sure that the electronic structure of GdFeCo is the same. We therefore raise the question on how factors such as the relative concentration of RE and TM \cite{liu2022giant}, concentration gradients \cite{cespedes2021current, zheng2021field}, or material aging \cite{sala2022asynchronous, sala2022ferrimagnetic} influence the resulting $\lambda_{SOC}$ sign. It also raises the question of the contribution of additional charge-to-spin conversion mechanisms, such as the Rashba effect, known to occur in Gd and capable of reversing sign upon oxidation \cite{krupin2005rashba}, as well as the possible involvement of orbital currents \cite{sala2022giant}. This is a matter for further investigation, which is beyond the scope of this study.

%often viewed as the spin analogue of the anomalous Hall effect (AHE), the latter being widely used to probe magnetisation. Although the microscopic origin of the AHE sign remains debated, being attributed variously to the RE sublattice \cite{ogawa1975reversal}, the TM sublattice \cite{mimura1976hall, asomoza1977extraordinary, malmhall1983extraordinary}, or to both sublattices \cite{park2021unconventional, hai2023novel}, it is well established that the AHE signal reverses sign at the magnetisation compensation point (as in Fig.~\ref{T charac}). One might therefore expect the SAHE symmetry to reverse as well. This naturally raises the question of why the symmetry of the SAHE does not undergo a similar sign reversal across compensation. 

\subsection{Discussion on the efficiencies amplitude}

Beyond the sign, the evolution of the amplitude of the torque efficiencies provides additional insight. The damping-like efficiency $\xi_{DL}$ associated with the SHE symmetry increases as the system approaches the magnetisation compensation temperature. Two possible mechanisms may account for this behaviour. First, a longer spin dephasing length near compensation would allow a larger fraction of the spin current generated inside the ferrimagnet to reach the NiFe layer. Such behaviour has been reported in related ferrimagnets such as GdCo \cite{lim2021dephasing} and CoTb multilayers \cite{yu2019long}. Alternatively, a small canting of the magnetisation away from the out-of-plane direction would increase the SAHE contribution when the temperature deviates from the magnetisation compensation, effectively decreasing the strength of the SHE contribution due to their opposite signs. Finally, the smaller DL efficiency observed in the Gd-dominant regime compared with the FeCo-dominant regime can be understood within the model of Fig.~\ref{fig:model}(b), in which the flow direction of the majority electrons reverses across the compensation temperature.

%\subsection{Discussion on self-torques sign change}

%The previously reported sign reversals of the damping-like self-torque in GdFeCo ferrimagnet \cite{cespedes2021current} across the magnetisation temperature cannot be attributed to a reversal of the SHE spin current itself, since our results show that its sign remains constant. Instead, these reversals are likely caused by variations in the spin-current absorption mechanism within the ferrimagnetic layer. One hypothesis is that this phenomena can be due to the gradient of concentration usually presents in such ferrimagnets \cite{damas2022ferrimagnet, guo2024controllable}.

\section{Conclusion}
In summary, we have used complementary ST-FMR techniques to disentangle the torques arising from spin currents of different symmetries generated by a GdFeCo ferrimagnet and to track their evolution across the magnetisation compensation temperature. By combining lineshape and DC-bias analyses, we isolated the torque components arising from the SHE and the SAHE and determined both their signs and their amplitude as functions of temperature. Our results show that the torques associated with the SHE symmetry retain a constant sign through compensation. In contrast, the SAHE contribution to the DL torque has the opposite sign, dominates over the SHE component in the parallel configuration, and also maintains its sign across both the magnetisation and angular compensation points. The opposite signs between the SHE- and SAHE-driven DL torque and the absence of any sign reversal across the compensations can be explained by considering that the SHE and SAHE originate from distinct electronic subsystems within GdFeCo that possess opposite SOC signs: the SHE is carried by the weakly polarised 5d electrons of Gd and is therefore insensitive to the magnetisation state, while the SAHE arises from the 3d polarised conduction electrons of FeCo, whose polarisation remains unchanged across the magnetisation and angular compensations. Furthermore, the enhancement of the SHE-driven DL torque near compensation suggests either an increased spin dephasing length or a slight deviation of the magnetisation from the out-of-plane direction, providing additional insight into the microscopic mechanisms governing spin transport in ferrimagnetic systems. The absence of any sign reversal in either spin current symmetry demonstrates that previously reported sign changes of the DL self-torque in GdFeCo can be driven by variations in spin current absorption within the ferrimagnet rather than by reversals of the spin currents themselves. Altogether, these results shed light on how the sublattices contribute to the spin current generation in TM-RE ferrimagnets, highlighting pathways to exploit different spin current symmetries for generating spin torques in adjacent layers or within the ferrimagnet itself.

\begin{widetext}

\begin{table}[h]
\centering
\begin{tabular}{|c|c|c||c|c|c|c|c|c|c|}
\hline
 \multicolumn{3}{|c||}{} & \multicolumn{2}{c|}{LS} & \multicolumn{4}{c|}{DC} \\ \hline
T & -dominant & config. &  $h_{DL}^{SHE}$ & $h_{FL'}^{SHE}$  & $\xi_{DL}^{SHE}$ & $\xi_{FL'}^{SHE}$&$\xi_{DL}^{SAHE+SHE}$  & $\xi_{FL'}^{SAHE+SHE}$  \rule{0pt}{1em} \\ \hline 

15K   & Gd&  $\parallel$   &$>0$       & $>0$      &  -& - &$-0.20 \pm 0.03$& $+0.04 \pm0.02$ \\ \hline
120K   &Gd & $\perp$         &$>0$   & $>0$         & $+0.05 \pm 0.06$ & $+0.03 \pm 0.04$& -& -  \\ \hline
130K    &Gd  & $\perp$       &$>0$  & $>0$    &   $+0.13 \pm0.03 $& unclear & - & -   \\ \hline
140K & Gd & $\perp$        &$>0$      &$>0$         &  $+0.13 \pm 0.03$ & $+0.03 \pm 0.02$&- & -     \\ \hline
150K   &FeCo  & $\perp$      &$>0$  & $>0$        & unclear & $+0.02 \pm 0.01$ & - & -                      \\ \hline
160K  &FeCo  & $\perp$     &$>0$    & $>0$        &  $+0.15 \pm 0.06$ & $+0.04 \pm 0.01$  & - & -  \\ \hline
300K   &FeCo  & $\parallel $  &$>0$       & $>0$           &  - & -& $-0.45\pm 0.07$ & $+0.10 \pm 0.01$ \\ \hline \hline
300K   &Pt  &    &  $>0$     & $>0$          &    $+0.14 \pm 0.01$ & $+0.05\pm 0.01$ & - & - \\ \hline
\end{tabular}
\caption{Sign of the spin torque related quantities in GdFeCo(10)/Cu(4)/NiFe(4) extracted by the lineshape analysis (LS) or by the DC-bias technique (DC). The spin torque related quantities in Pt(5)/NiFe(4) reference sample are also displayed.}
\label{Table_SAHE+SHE}
\end{table}

\end{widetext}

\begin{acknowledgments}
%We thanks funds by the French National Research Agency (ANR) through project ANR-19-CE24-0016-01 ‘Toptronic ANR’.
This work was funded by the ERC CoG project MAGNETALLIEN grant ID  101086807, the EU-H2020-RISE project Ultra Thin Magneto Thermal Sensoring ULTIMATE-I (Grant ID. 101007825), and the French National Research Agency (ANR) through the project ANR-19-CE24-0016-01 ‘Toptronic ANR’, “Lorraine Université d’Excellence” reference ANR-15-IDEX-04-LUE. It was also partially supported by the ANR through the France 2030 government grants EMCOM (ANR-22-PEEL-0009), PEPR SPIN ANR-22-EXSP-0007 and ANR-22-EXSP-0009.
Devices in the present study were patterned at Institut Jean Lamour's clean room facilities (MiNaLor). These facilities are partially funded by FEDER and Grand Est region.% through the RANGE project
\end{acknowledgments}

\bibliography{Main_Manuscript}

\end{document}